\documentstyle[12pt]{article}
\begin{document}

\
\vskip 4truecm
\centerline{{\bf NON-EQUILIBRIUM QUANTUM FIELD EVOLUTION}}
\centerline{{\bf IN FRW COSMOLOGIES }}

\bigskip

\bigskip

\bigskip

\centerline{{\bf H.J. de Vega}}

\bigskip

\hspace*{-6mm}{\it {Laboratoire de Physique Th\'eorique et Hautes
Energies, Universit\'e Pierre et Marie Curie (Paris VI) ,
Laboratoire Associ\'{e} au CNRS UA280.
Tour 16, 1er. \'etage, 4, Place Jussieu
75252 Paris, Cedex 05, France\\ }}

\bigskip

\bigskip

\bigskip

\centerline{{\bf {\it Based on a lecture at the NATO Advanced Research
Workshop on}}}
 \centerline{{\bf{\it ELECTROWEAK PHYSICS AND THE EARLY UNIVERSE}}}
 \centerline{{\bf{\it held at Sintra, Portugal from March 22 to 27, 1994.}}}

\pagebreak
\
\vskip 8truecm
\centerline{{\bf NON-EQUILIBRIUM QUANTUM FIELD EVOLUTION}}
\centerline{{\bf IN FRW COSMOLOGIES }}

\bigskip

\bigskip

\bigskip

\centerline{{\bf H.J. de Vega}}

\bigskip

\hspace*{-6mm}{\it {Laboratoire de Physique Th\'eorique et Hautes
Energies, Universit\'e Pierre et Marie Curie (Paris VI) ,
Laboratoire Associ\'{e} au CNRS UA280.
Tour 16, 1er. \'etage, 4, Place Jussieu
75252 Paris, Cedex 05, France\\ }}

\bigskip

\bigskip

We derive the effective equations for the out of equilibrium time evolution of
the order parameter and the fluctuations of a scalar field theory in spatially
flat FRW cosmologies.
After setting the problem in general we propose a
non-perturbative, self-consistent Hartree approximation. The method consists of
evolving an initial functional thermal density matrix in time and is suitable
for studying phase transitions out of equilibrium.  The renormalization aspects
are studied in detail and we find that the counterterms depend on the initial
state. We investigate the high temperature expansion and show that it breaks
down at long times.   The infinite time limit is computed for de
Sitter spacetime. We obtain the time evolution of the initial Boltzmann
distribution functions, and argue that in the Hartree
approximation, the time evolved state is a ``squeezed'' state. We illustrate
the departure from thermal equilibrium by  studying the case of a
free  field in de Sitter and radiation dominated cosmologies. It
is found that a suitably defined non-equilibrium entropy per mode increases
linearly with comoving time in a de Sitter cosmology, whereas it is {\it not} a
monotonically increasing function in the  radiation dominated case.
This work has been done in collaboration with D. Boyanovsky and R. Holman.

\section{\bf Introduction}

 The usual assumption in inflation  \cite{infla} is  that the dynamics of the
spatial zero mode of the (so-called) inflaton field is governed by some
approximation by the effective potential which incorporates the effects of
quantum fluctuations of the field. Thus the equation of motion is
usually an ordinary  non-linear differential equation of the form:

\begin{equation}
\ddot{\phi} + 3\frac{\dot{a}}{a}\dot{\phi} + V'_{\rm eff}(\phi) = 0
\end{equation}

where $a(t)$ is the expansion factor in the metric:
\begin{equation}
ds^2 = dt^2-a^2(t)~d\vec{x}^2
\label{met}
\end{equation}
The problem here is that the effective potential is really only suited for
analyzing {\em static} situations; it is the effective action evaluated for a
field configuration that is constant in time \cite{effecpot}. Thus, it
is inconsistent to use the effective potential in a {\em dynamical}
situation. Notice that such inconsistency appears for {\bf any}
inflationary scenario (old, new, chaotic, ...).

More generally, the standard methods of high temperature field theory are
based on an equilibrium formalism\cite{dolan,sweinberg}; there is no time
evolution in such a
situation. Such techniques preclude us from treating non-equilibrium
situations
such as surely exist for very weakly coupled theories in the early universe.

More precisely, the effective potential develops an
imaginary part around points
$\phi_0$ where $ V''(\phi_0) < 0 $ [$\phi_0$ : false vacuum]. This
fact leads to long-wavelength modes with imaginary frequencies $
\omega(k) = \sqrt{ k^2 + V''(\phi_0) } $ for $ 0 \leq k <
\sqrt{ |V''(\phi_0)| } $. When the field $\phi$ starts localized
around $\phi_0$ , these {\bf long-wavelength} unstable  modes push the
system to evolve on time away from  $\phi_0$. The smaller is the
non-linear coupling, the better is this small fluctuations
approximations and the more important are these unstabilities.

In collaboration with Daniel  Boyanovsky and Rich Holman
we try  to rectify this situation by addressing three
issues\cite{bd,bds,bdh,bis}:
a) obtaining the evolution equations for the order parameter including
the {\it quantum fluctuations},
b) studying departures from thermal equilibrium if
the initial state is specified as a thermal ensemble, c) understanding the
renormalization aspects and the validity of the high temperature expansion.

Our ultimate goal is to study the dynamics of phase transitions in the early
universe, in particular, the formation and evolution of correlated domains and
symmetry breaking in an expanding universe.  From some of our previous studies
on the dynamics of phase transitions\cite{bd,bds} in Minkowski space, we
have learned that the familiar picture of ``rolling'' is drastically modified
when the fluctuations are taken into account. As the phase transition proceeds
fluctuations become large and correlated regions (domains) begin to grow. This
enhancement of the fluctuations modifies substantially the evolution equation
of the order parameter. Thus the time dependence of the order parameter is not
enough to understand the dynamical aspects of the phase transition; it must
be studied in conjunction with that of the fluctuations.

Our approach is to use the functional Schr\"{o}dinger formulation, wherein we
specify the initial wavefunctional $\Psi[\Phi(\vec{.}); t]$
(or more generally a
density matrix $\rho[\Phi(\vec{.}), \tilde{\Phi}(\vec{.}); t]$), and then use
the Schr\"{o}dinger equation to evolve this state in time. We can then use
this state to compute all of the expectation values required in the
construction of the effective equations of motion for the order parameter of
the theory, as well as that for the fluctuations.

One advantage of this approach is that it is truly a dynamical one; we set up
initial conditions at some time $t_o$ by specifying the initial state and then
we follow the evolution of the order parameter $\phi(t) \equiv
\langle\Phi(\vec{x})\rangle$ and of the fluctuations as this state evolves in
time. Another advantage is that it allows for departures from equilibrium.
Thus, issues concerning the restoration of symmetries in the early universe can
be addressed in a much more general setting.

Our analysis applies quite generally to any arbitrary spatially flat FRW
cosmology.
We also determine the time evolution of
the initial
(Boltzmann) distribution functions, relate the time evolution to ``squeezed
states'' and perform a numerical integration in the case of free fields for de
Sitter and radiation dominated cosmologies. We expect to provide a quantitative
analysis of the evolution of the order parameter and the dynamics
of  phase transitions for interacting fields in a forthcoming article
\cite{bis}.

The initial state we pick for the field $\Phi(\vec{x},t)$ is that corresponding
to a thermal density matrix centered at $\phi(t)$. It is then useful to try to
understand the high temperature limit of our calculations. We are able to
compute both the leading and subleading terms in the high $T$ expansion of
$\langle \phi^2(t)\rangle$.
{}From this we show that the high $T$ expansion cannot
be valid for all time, but breaks down in the large time limit.
We then compute the limit of long  times ($t \to \infty$).
during the phase transition.

The time evolution of the Boltzmann distribution functions (initially the
thermal equilibrium distribution functions) is obtained in
ref.\cite{bdh}. It is pointed out
that to one-loop order and also in the Hartree approximation, the time
evolved density matrix describes quantum ``squeezed'' states and time
evolution corresponds to a Bogoliubov transformation.

To illustrate the departure of equilibrium, we have studied
numerically in ref.\cite{bdh} the
case of a free massive scalar field in de Sitter and radiation dominated
cosmologies. It was found that a suitably defined coarse-grained
non-equilibrium entropy (per $\vec{k}$ mode) grows linearly with time in
the de Sitter case but it is not a monotonically increasing function of
time in the radiation dominated case. This result may cast some doubt on the
applicability of this definition of the non-equilibrium entropy.
There still remain some (open) fundamental questions regarding the
connection of this entropy and the thermodynamic entropy of the universe,
in particular whether the amount of entropy produced is consistent with
the current bounds.

This work sets the stage for a numerical study of the dynamics of phase
transitions in cosmology fully incorporating the non-equilibrium aspects in the
evolution of the order parameter and which at the same time can account for the
dynamics of  the fluctuations which will necesarily become very important
during the phase transition.

We have applied similar methods to investigate the formation of
disordered chiral condensates in high energy collisions \cite{quir}.

\section{\bf Evolution Equations}

We consider the inflaton scalar field $\Phi(\vec{x},t)$ in
the  spatially flat FRW cosmology (\ref{met}) with
action given by
\begin{eqnarray}
S         & =  & \int d^4x
 a^3(t)\left[\;\frac{1}{2}\dot{\Phi}^2(\vec{x},t)-\frac{1}{2}
\frac{(\vec{\nabla}\Phi(\vec{x},t))^2}{a(t)^2}-V(\Phi(\vec{x},t))\;\right]
 \label{action} \\
V(\Phi)   & =  & \frac{1}{2}[m^2+ \xi {\cal{R}}] ~\Phi^2(\vec{x},t)+
\frac{\lambda}{4!}~\Phi^4(\vec{x},t) \label{potential} \\
{\cal{R}}    & =  & 6\left(\frac{\ddot{a}}{a}+\frac{\dot{a}^2}{a^2}\right)
\end{eqnarray}
with ${\cal{R}}$ the Ricci scalar and $\xi$ a non-minimal coupling constant.

In the Schr\"{o}dinger representation (at an arbitrary fixed time
$t$), the Hamiltonian becomes
\begin{equation}
H(t) = \int d^3x \left\{ -\frac{\hbar^2}{2a^3(t)}
\frac{\delta^2}{\delta \Phi(\vec{x})^2}+
\frac{a(t)}{2}(\vec{\nabla}\Phi)^2+
a^3(t) ~V(\Phi) \right\} \label{hamiltonian}
\end{equation}

Since we  consider a `thermal ensemble', we
work with a (functional) density matrix $\hat{\rho}$
with matrix elements in the
Schr\"{o}dinger representation
$\rho[\Phi(\vec{.}), \tilde{\Phi}(\vec{.});t]$. We will
{\it assume} that the
density matrix obeys the functional Liouville equation
\begin{equation}
i\hbar \frac{\partial \hat{\rho}}{\partial t} = \left[H(t),\hat{\rho}\,\right]
\label{liouville}
\end{equation}

 Normalizing the density matrix such that $Tr\hat{\rho}=1$, we define
as `order parameter'
\begin{equation}
\phi(t) = \frac{1}{\Omega}\int d^3x \langle \Phi(\vec{x},t) \rangle =
\frac{1}{\Omega}\int d^3x \; Tr\left[\hat{\rho}(t)\Phi(\vec{x})\right]
\label{orderparameter}
\end{equation}
where $\Omega$ is the comoving volume.

The evolution equations for the order parameter are as follows:
\begin{equation}
\frac{d^2 \phi(t)}{dt^2}  + \frac{3}{a(t)}\frac{da(t)}{dt} +
 \frac{1}{\Omega}\int d^3x  ~ \langle
 \frac{\delta V(\Phi)}{\delta \Phi(\vec{x})} \rangle = 0 \label{evo}
\end{equation}
It is now convenient to write the field in the {Schr\"{o}dinger} picture as
\begin{eqnarray}
\Phi(\vec{x})   & = & \phi(t)+\eta(\vec{x},t) \label{split} \\
\langle \eta(\vec{x},t) \rangle
     & = & 0 \label{doteta}
\end{eqnarray}

Expanding the right hand side of (\ref{evo}) we find the effective
equation of motion for the order parameter:
\begin{equation}
\frac{d^2 \phi(t)}{dt^2}+3 \, \frac{\dot{a}(t)}{a(t)} \,
\frac{d \phi(t)}{dt}+V'(\phi(t))+\frac{V'''(\phi(t))}{2 \Omega}\int d^3x~
 \langle \eta^2(\vec{x},t)\rangle+\cdots
=0 \label{effequation}
\end{equation}
where primes stand for derivatives with respect to $\phi$.

This equation of motion is clearly {\em very} different from the one obtained
by using the effective potential. It may be easily seen (by writing the
effective action as the classical action plus the logarithm of the determinant
of the quadratic fluctuation operator) that this is the equation of motion
obtained by the variation of the one-loop effective action.

The {\it static} effective potential is clearly not the appropriate quantity to
use to describe scalar field dynamics in an expanding universe. Although there
may be some time regime in which the time evolution is slow and fluctuations
rather small, this will certainly {\it not} be the case at the onset of a phase
transition. As the phase transition takes place, fluctuations become dominant
and grow in time signaling the onset of long range
correlations\cite{bd,bds}.

\section{\bf Hartree equations}

Motivated by our previous studies in Minkowski space\cite{bd,bds} which
showed that the growth of correlation and enhancement of fluctuations during
a phase transition may not be described perturbatively, we proceeded to
obtaining the equations of motion in a Hartree approximation\cite{bdh}. This
approximation is non-perturbative in the sense that it sums up infinitely
many diagrams of the cactus or bubble-type\cite{dolan}.
The Hartree approximation
becomes exact in the $N \rightarrow \infty$ limit of an $O(N)$ vector theory.
It at least
provides a consistent non-perturbative framework in which correlations
and fluctuations can be studied.
Corrections to the Hartree approximation
can be systematically implemented in a consistent fashion.

The Hartree self-consistent approximation is implemented as follows. We
decompose the field as in (\ref{split}), using a potential as in
(\ref{potential}). We find that the Hamiltonian becomes
\begin{eqnarray}
H & = &  \int d^3x \left\{-\frac{\hbar^2}{2 a^3(t)}\frac{\delta^2}
{\delta\eta^2}+\frac{a(t)}{2}\left(\vec{\nabla}\eta\right)^2+a^3(t)
\left(V(\phi)+V'(\phi)\;\eta+\frac{1}{2!}V''(\phi)\; \eta^2 \right. \right.
\nonumber \\
  & + &
\left. \left. \frac{1}{3!} \lambda \, \phi\, \eta^3+
\frac{1}{4!} \lambda \, \eta^4 \right)\right\} \label{fullham}
\end{eqnarray}

The Hartree approximation is obtained by assuming the factorization
\begin{eqnarray}
\eta^3 (\vec{x},t) &  \rightarrow & 3 \langle \eta^2(\vec{x},t) \rangle
\eta(\vec{x},t) \label{eta3} \\
\eta^4 (\vec{x},t) &  \rightarrow & 6 \langle \eta^2(\vec{x},t) \rangle
\eta^2(\vec{x},t) -3
\langle \eta^2(\vec{x},t) \rangle^2 \label{eta4}
\end{eqnarray}
where $\langle \cdots \rangle$ is the average using the time evolved density
matrix. This average will be determined self-consistently (see below).
Translational invariance shows that $\langle \eta^2(\vec{x},t) \rangle$ can
only be a function of time. This approximation makes the Hamiltonian quadratic
at the expense of a self-consistent condition. In the time independent
(Minkowski) case this approximation sums up all the daisy (or cactus)
diagrams and leads to the self-consistent gap equation\cite{dolan}. In this
approximation the Hamiltonian becomes
\begin{eqnarray}
& & H = \Omega ~a^3(t) ~ {\cal{V}}(\phi) +  \label{hartham} \\
& &  \int d^3x \left\{-\frac{\hbar^2}{2\; a^3(t)}\; \frac{\delta^2}{\delta
\eta^2}+ \frac{a(t)}{2}\left(\vec{\nabla}\eta\right)^2
+a^3(t)\left({\cal{V}}^{(1)}\,(\phi)\eta+\frac{1}{2}
{\cal{V}}^{(2)}\,(\phi)\eta^2 \right)
\right\} \nonumber \\
& & {\cal{V}}(\phi) = V(\phi)- \frac{1}{8}\lambda \langle \eta^2 \rangle^2
 \; , \;
{\cal{V}}^{(1)}(\phi) = V'(\phi)+\frac{1}{2}\lambda\;\phi\;
\langle \eta^2 \rangle   \; , \;
{\cal{V}}^{(2)}(\phi) = V''(\phi)+ \frac{1}{2}
\lambda  \langle \;\eta^2 \rangle ~.\nonumber
\end{eqnarray}

It is convenient to introduce the discrete Fourier transform of the
fields in the comoving frame as
\begin{equation}
\eta(\vec{x},t) = \frac{1}{\sqrt{\Omega}}\sum_{\vec{k}}
\eta_{\vec{k}}(t) ~ e^{-i\vec{k}\cdot\vec{x}}
\label{fourier1}
\end{equation}
The Hamiltonian (\ref{hartham}) then  takes the form
\begin{eqnarray}
H & = & \Omega \; a^3(t) \; V(\phi(t))+\nonumber \\
 &   & \frac{1}{2} \sum_{\vec{k}}
\left\{ -\frac{\hbar^2}{a^3(t)}
 \frac{\delta^2}{\delta \eta_{\vec{k}}\delta\eta_{-\vec{k}}}+2\,a^3(t)\;
 {\cal V}'_{\vec{k}}(\phi(t))\;\eta_{-\vec{k}}+
\omega^2_k(t)\;\eta_{\vec{k}}\;\eta_{-\vec{k}}\right\} \label{hamodes}
\end{eqnarray}
where the time dependent frequencies ($\omega^2_k(t)$)
and the linear term in $\eta$ have the values
\begin{eqnarray}
\omega_k^2(t) & = & a(t) \;\vec{k}^2+a^3(t)\;{\cal{V}}^{(2)}(\phi(t))
 \label{newfreq} \\
{\cal{V}}^{(1)}_{\vec{k}}(\phi(t))
     & = & {\cal{V}}^{(1)}(\phi(t))
\;\sqrt{\Omega}\; \delta_{\vec{k},0}
\end{eqnarray}
We propose the following Gaussian ansatz for the functional density
matrix elements in the {Schr\"{o}dinger} representation
\begin{eqnarray}
\rho[\Phi,\tilde{\Phi},t]&=&\prod_{\vec{k}} {\cal{N}}_k(t) \exp\left\{
- \left[\frac{A_k(t)}{2\hbar}\eta_k(t)\eta_{-k}(t)+
\frac{A^*_k(t)}{2\hbar}\tilde{\eta}_k(t)\tilde{\eta}_{-k}(t)+
\frac{B_k(t)}{\hbar}\eta_k(t)\tilde{\eta}_{-k}(t)\right] \right. \nonumber \\
       &   & \left. +\frac{i}{\hbar}\;\pi_k(t)\;\left(\eta_{-k}(t)-
\tilde{\eta}_{-k}(t)\right) \right\} \label{densitymatrix} \\
\eta_k(t)          & = & \Phi_k-\phi(t)\sqrt{\Omega}\;\delta_{\vec{k},0}
\quad . \quad
\tilde{\eta}_k(t) = {\tilde{\Phi}}_k -\phi(t)\sqrt{\Omega}\;\delta_{\vec{k},0}
\nonumber
\end{eqnarray}
where $\phi(t) = \langle \Phi(\vec{x}) \rangle$ and $\pi_k(t)$ is the Fourier
transform of $\langle \Pi(\vec{x}) \rangle$. This form of the density matrix
is dictated by the hermiticity condition $\rho^{\dagger}[\Phi,\tilde{\Phi},t] =
\rho^*[\tilde{\Phi},\Phi,t]$; as a result of this, $B_k(t)$ is real.
The kernel $B_k(t)$ determines the amount of ``mixing'' in the
density matrix, since if $B_k=0$, the density matrix corresponds to a pure
state because it is a wave functional times its complex conjugate.

In order to solve for the time evolution of the density matrix
(\ref{liouville}) we need to specify the density matrix at some initial
time $t_o$. It is at this point that we have to {\it assume} some physically
motivated initial condition. We believe that this is a subtle point that
has not received proper consideration in the literature. A system in
thermal equilibrium has time-independent ensemble averages (as the evolution
Hamiltonian commutes with the density matrix) and there is no memory of any
initial state. However, in a time dependent background, the density matrix
will evolve in time, departing from the equilibrium state and
correlation functions or expectation values may depend on
details of the initial state.

We will {\it assume} that at early times
the initial density matrix is {\em thermal} for the modes that
diagonalize the
Hamiltonian at $t_o$ (we call these the {\em adiabatic} modes). The effective
temperature for these modes is $k_BT_o = 1/\beta_o$. It is only in this
initial state that the notion of ``temperature'' is meaningful. As the
system departs from equilibrium one cannot define a thermodynamic temperature.
Thus in this case the ``temperature'' refers to the temperature defined in the
initial state.

 Inserting the Gaussian Ansatz (\ref{densitymatrix}) into the Liouville
equation yields upon using the hamiltonian (\ref{hartham}) the
following equations for the functions $ A_k(t) , B_k(t) $ and ${\cal
N}_k(t)$:
\begin{eqnarray}
i\frac{\dot{{\cal{N}}}_k}{{\cal{N}}_k} & = & \frac{1}{2a^3(t)}(A_k-A^*_k)
\label{normeq} \\
i\dot{A}_k                     & = &
\left[ \frac{A^2_k-B^2_k}{a(t)^3}-\omega^2_k(t)\right] \label{Aeq} \\
i \dot{B_k}                    & = & \frac{B_k}{a^3(t)}(A_k-A^*_k)
\label{Beq}
\end{eqnarray}
The equation for $B_k(t)$ reflects the fact that a pure state
$B_k=0$ remains pure under time evolution.

Writing $A_k$ in terms of its real and imaginary components
$A_k(t) = A_{Rk}(t)+i A_{Ik}(t)$ (and because $B_k$ is real) we
find that
\begin{equation}
\frac{B_k(t)}{A_{Rk}(t)} = \frac{B_k(t_o)}{A_{Rk}(t_o)}
\label{invar}
\end{equation}
and that  the time evolution is unitary (as it should be), that is
\begin{equation}
\frac{{\cal{N}}_k(t)}{\sqrt{\left(A_{Rk}(t)+B_k(t)\right)}} = \mbox{constant}
\label{unitarity}
\end{equation}
For the chosen thermal initial conditions at $ t = t_0 $, we have
\begin{eqnarray}
A_k(t_o) & = & A^*_k(t_o) =
  {\cal{W}}_k(t_o)~a^3(t_o)~\coth\left[\beta_o\hbar  {\cal{W}}_k(t_o)
\right] \label{Ato} \\
B_k(t_o) & = & - \frac{ {\cal{W}}_k(t_o)a^3(t_o)}
{\sinh\left[\beta_o\hbar  {\cal{W}}_k(t_o)\right]}
\nonumber \\
{\cal{N}}_k(t_o)
   & = &  \left[\frac{ {\cal{W}}_k(t_o)a^3(t_o)}{\pi\hbar}\tanh\left[
\frac{\beta_o\hbar  {\cal{W}}_k(t_o)}{2}\right]\right]^{\frac{1}{2}}
 \label{normo}
\end{eqnarray}
where
 \begin{equation}
 {\cal{W}}_k(t_o)  =  {{\left[ {\vec{k}^2}+
{m^2(T_o)} \right]^{\frac{1}{2}}}\over {a(t_o)}}
\qquad ,\qquad
\frac{m^2(T_o)}{a^2(t_o)}               =  {\cal{V}}^{(2)}(\phi(t_o))
\label{defmass}
\end{equation}
 The initial density matrix is normalized such that
$Tr\rho(t_o)=1$. Since time evolution is unitary such a normalization
will be constant in time. For $T_o = 0$ the density matrix describes a
pure state since $B_k = 0$.

It is convenient to introduce the complex function:
\begin{equation}
 {\cal{A}}_k(t)=\tanh\left[\beta_o\hbar W_k(t_o)\right] {\rm Re} A_{k}(t)
+ i  ~ {\rm Im}  A_{k}(t)
\end{equation}
Then  ${\cal{A}}_k(t)$ obeys a   Riccati equation
\begin{equation}
i\dot{\cal{A}}_k(t) = \frac{1}{a^3(t)}\left[{{\cal{A}}_k}^2(t)-
\omega^2_k(t)a^3(t) \right] \label{riccati}
\end{equation}
with the initial conditions:
\begin{equation}
{\cal{A}}_{k}(t_o)  =   {\cal{W}}_k(t_o)a^3(t_o) \label{Are}
\end{equation}
 Eq.(\ref{riccati}) can be easily linearized by introducing
the functions $\varphi^H_k(t)$ as
\begin{equation}
{\cal{A}}_k(t) = -ia^3(t)\frac{\dot{\varphi}^H_k(t)}{\varphi^H_k(t)}
\label{changeofvar}
\end{equation}
 The equal time two-point function thus becomes
\begin{equation}
\langle \eta^2(\vec{x},t) \rangle =  \frac{\hbar}{2}
 \int \frac{d^3k}{(2\pi)^3}
\mid \varphi^H_k(t) \mid^2
\coth\left[\beta_o\hbar {\cal{W}}_k(t_o)/2
\right], \label{newtwopointfunc}
\end{equation}
which leads to the following set of self-consistent time dependent
Hartree equations:
\begin{eqnarray}
& & \ddot{\phi}+3 \frac{\dot{a}}{a}\dot{\phi}+V'(\phi)+
\lambda \, \phi \, \frac{\hbar}{2}  \int \frac{d^3k}{(2\pi)^3}
\frac{\mid \varphi^H_k(t) \mid^2}{2}
\coth\left[\beta_o\hbar {\cal{W}}_k(t_o)/2
\right] = 0 \label{harteqofmot} \\
& & \left[\frac{d^2}{dt^2}\,+\,3 \,\frac{\dot{a}(t)}{a(t)}\frac{d}{dt}+
\frac{\vec{k}^2}{a^2(t)}\,+\,V''(\phi(t)) \; + \right. \nonumber \\
& & \left. \lambda \,\frac{\hbar}{2}
 \int \frac{d^3k}{(2\pi)^3}
\frac{\mid \varphi^H_k(t) \mid^2}{2}
\coth\left[\beta_o\hbar {\cal{W}}_k(t_o)/2
\right] \; \right]
\varphi^H_k(t)
=0 \label{newdiffeqU} \\
& &
\varphi^H_k(t_o)        =  \frac{1}{\sqrt{a^3(t_o){\cal{W}}_k(t_o)}}
\quad , \quad
\dot{\varphi}^H_k(t)\mid_{t_o}  =
 i\sqrt{\frac{{\cal{W}}_k(t_o)}{a^3(t_o)}}
\label{newboundconUdot}
\end{eqnarray}
 That is, we are faced with two complicated coupled non-linear and non-local
equations for $\phi(t)$ and $\varphi_k(t)$.

The Hartree self-consistent equations (\ref{harteqofmot}) need
regularization and
renormalization since the momentum
integrals of the mode functions (\ref{newtwopointfunc}) diverge.

Because the Bose-Einstein distribution
functions are exponentially suppressed at large momenta, the finite
temperature contribution will always be convergent and we need only address
the zero temperature contribution.

A WKB analysis of eq.(\ref{harteqofmot}) provides the large $k$
behaviour of the mode functions $\varphi_k(t) $ \cite{bdh}:
 \begin{equation}
\frac{\mid \varphi^H_k(t) \mid^2}{2} \buildrel{ k\to \infty}\over=
 \frac{1}{2a^2(t)k}+\frac{1}{4k^3}
\left[\frac{\dot{a}^2(t_o)}{a^2(t)}-\left(-\frac{{\cal{R}}}{6}+V''(\phi)+
\frac{\lambda}{2} \langle \eta^2(\vec{x},t) \rangle \right)\right]+
 {\cal{O}}(1/k^4)+\cdots \label{divin}
\end{equation}
 Introducing an upper momentum cut-off $\Lambda$ we obtain
\begin{equation}
\langle \eta^2 (\vec{x},t) \rangle
= \frac{\hbar}{8\pi^2} \frac{\Lambda^2}{a^2(t)}+
 \frac{\hbar}{8\pi^2}
\ln \left(\frac{\Lambda}{K}\right)
 \left[\frac{\dot{a}^2(t_o)}{a^2(t)}-\left(-\frac{{\cal{R}}}{6}+
V''(\phi(t))+\frac{\lambda}{2} \langle \eta^2(\vec{x},t) \rangle
\right)\right]+ \mbox{ finite } \label{divergences}
\end{equation}
where we have introduced a renormalization point $K$, and the finite part
depends on time, temperature and $K$.

There are several physically important features of the divergent structure
obtained above. First, the quadratically divergent term reflects the fact that
the physical momentum cut-off is being red-shifted by the expansion. (This term
will not appear in dimensional regularization).

Secondly, the logarithmic divergence contains a term that reflects the initial
condition (the derivative of the expansion factor at the initial time $t_o$).
The initial condition breaks any remnant symmetry. For example, in de Sitter
space there is still invariance under the de Sitter group, but this is also
broken by the initial condition at an arbitrary time $t_o$.Thus this term is
not forbidden, and its appearance does not come as a surprise. As a consequence
of this term, we need a time dependent term in the bare mass proportional to
$1/{a^2(t)}$.

We are now in a position to present the renormalization prescription within
the Hartree approximation. The differential
equation for the mode functions (\ref{newdiffeqU})  {\it must be finite}.
Thus the renormalization conditions are obtained from
\begin{equation}
m^2_B(t) + \frac{\lambda_B}{2}\phi^2(t)+\xi_B{\cal{R}}+\frac{\lambda_B}{2}
\langle \eta^2 \rangle_B =
m^2_R+\frac{\lambda_R}{2}\phi^2(t)+\xi_R{\cal{R}}+\frac{\lambda_R}{2}
\langle \eta^2 \rangle_R \label{renormcond}
\end{equation}
where the subscripts $B,\ R$ refer to bare and
 renormalized quantities respectively and
$\langle \eta^2 \rangle_B$ is read from
 (\ref{divergences})
\begin{eqnarray}
& & \langle \eta^2 \rangle_B  =  \frac{\hbar}{8\pi^2}
\frac{\Lambda^2}{a^2(t)} \label{eta2} \\
 &  & + \frac{\hbar}{8\pi^2} \ln \left(\frac{\Lambda}{K}\right)
 \left[\frac{\dot{a}^2(t_o)}{a^2(t)}-\left(-\frac{{\cal{R}}}{6}+
m^2_R+\frac{\lambda_R}{2}\phi^2(t)+\xi_R{\cal{R}}+\frac{\lambda_R}{2}
\langle \eta^2 \rangle_R \right)\right]+ \mbox{ finite }
\nonumber
\end{eqnarray}
Using the renormalization conditions (\ref{renormcond}) we obtain
\begin{eqnarray}
& & m^2_B(t) +\frac{\lambda_B\hbar}{16\pi^2}\frac{\Lambda^2}{a^2(t)}+
\frac{\lambda_B\hbar}{16\pi^2}\ln \left(\frac{\Lambda}{K}\right)
 \frac{\dot{a}^2(t_o)}{a^2(t)}  =
m^2_R\left[1+\frac{\lambda_B\hbar}{16\pi^2}
\ln \left(\frac{\Lambda}{K}\right)\right] \label{massren} \\
& & \lambda_B                       =
\frac{\lambda_R}{1-\frac{\lambda_R\hbar}{16\pi^2}
\ln \left(\frac{\Lambda}{K}\right)} \label{lambdaren} \\
& & \xi_B                         =
\xi_R + \frac{\lambda_B\hbar}{16\pi^2}
\ln \left(\frac{\Lambda}{K}\right) \left(\xi_R-\frac{1}{6}\right)
\label{xiren} \\
& & \langle \eta^2 \rangle_R        =   I_R + J
\label{etaren}
\end{eqnarray}
where
\begin{eqnarray}
I_R & = & \hbar\int \frac{d^3k}{(2\pi)^3} \left\{
\frac{\mid \varphi^H_k(t) \mid^2}{2} -
\frac{1}{2ka^2(t)} + \right. \nonumber \\
 &   & \left. \frac{\theta(k-K)}{4k^3}\left[
-\frac{{\cal{R}}}{6}- \frac{\dot{a}^2(t_o)}{a^2(t)}
+ m^2_R+\frac{\lambda_R}{2}\phi^2(t)+\xi_R\;{\cal{R}}+\frac{\lambda_R}{2}
\langle \eta^2 \rangle_R
\right] \right\}
\label{irren} \\
J   & = & \hbar\int \frac{d^3k}{(2\pi)^3}
\frac{\mid \varphi^H_k(t)\mid^2}{\exp{\beta_o\hbar {\cal{W}}_k(t_o)} - 1}
\label{jota}
\end{eqnarray}

The conformal coupling $\xi = 1 / 6$ is a {\it fixed point} under
renormalization\cite{birrell}. In dimensional regularization the terms
involving $\Lambda^2$ are absent and $\ln\Lambda$ is replaced by a simple pole
at the physical dimension. Even in such a regularization scheme, however, a
time dependent bare mass is needed. The presence of this new renormalization
allows us to introduce a new renormalized mass term  of the form \[
\frac{\Sigma}{a^2(t)} \] This counterterm may be interpreted as a squared mass
red-shifted by the expansion of the universe. However, we shall set $\Sigma =
0$ for simplicity.

Notice that there is a weak cut-off dependence on the effective equation
of motion for the order parameter.

 For {\it fixed} $\lambda_R$, as the cutoff $\Lambda \rightarrow \infty$
\begin{equation}
  \lambda_B \approx -\frac{(4\pi)^2}{\ln\left(\frac{\Lambda}{K}\right)}
{}~,~~ \xi_B = \frac{1}{6} + O\left(\frac{1}{\ln\Lambda}\right)
{}~,~~  m^2_B(t) = \frac{1}{a^2(t)}\left[\frac{\Lambda^2}{\ln
\left(\frac{\Lambda}{K}\right)} +  {\dot{a}}^2(t_0)  \right]+
O\left(\frac{1}{\ln\Lambda}\right)
\label{mbaretime}
\end{equation}

This approach to $0^-$ of the bare coupling as the cutoff is removed
translates into an instability in the bare theory. This is a consequence of the
fact that the N-component $\Phi^4$ theory for $N \to \infty$ is asymptotically
free (see ref.\cite{asympto}), which is not relieved in curved space-time.
Clearly this theory is sensible only as a low-energy cut-off effective theory,
and it is in this restricted sense that we will ignore the weak cut-off
dependence and neglect the term proportional to the bare coupling in
(\ref{rennewdiffeqU}).

The renormalized self-consistent Hartree equations thus become
after letting $\Lambda = \infty$:
\begin{eqnarray}
& & \ddot{\phi}+3 \;\frac{\dot{a}}{a}\;\dot{\phi}+m^2_R\phi+\xi_R\;{\cal{R}}
\;\phi+ \frac{\lambda_R}{2} \;\phi^3+
\frac{\lambda_R \;\phi}{2} \;\langle \eta^2 \rangle_R =0
 \label{rennewdiffeqU} \\
& & \left[\;\frac{d^2}{dt^2}+3 \;\frac{\dot{a}(t)}{a(t)}\frac\;{d}{dt}+
\frac{\vec{k}^2}{a^2(t)}+m^2_R+ \xi_R \;{\cal{R}}+ \frac{\lambda_R}{2}
\;\phi^2 +
 \frac{\lambda_R}{2} \;\langle \eta^2 \rangle_R \;\right]
 \varphi^H_k(t) =0 \nonumber
\end{eqnarray}
where $\langle \eta^2 \rangle_R$ is given by equations (\ref{irren}),
(\ref{jota}).

\section{High Temperature Limit}

One of the payoffs of understanding the large-$k$ behavior of the mode
functions (as obtained in the previous section via the WKB method) is
that it permits the evaluation of the high temperature limit. We shall
perform our analysis of the high temperature expansion for the Hartree
approximation.

The finite temperature contribution is determined by the integral

\begin{equation}
J =
\hbar
\int \frac{d^3k}{(2\pi)^3}
\frac{\mid \varphi^H_k(t) \mid^2}{e^{\beta_o
\hbar {\cal{W}}_k(t_o)}-1}  \label{tempint}
\end{equation}
For large temperature, only momenta $k \geq T_o$ contribute. Thus the
leading contribution is determined by the first term in
eq.(\ref{divin})). We find
\begin{equation}
J = \frac{1}{12 \hbar}
\left[ \frac{k_B T_o a(t_o)}{a(t)}\right]^2
\left[1+{\cal{O}}(1/T_o) + \cdots \right] \label{highT}
\end{equation}

Thus we see that the leading high temperature behavior results in an
effective time dependent temperature

\[ T_{eff}(t) = T_o \left[\frac{a(t_o)}{a(t)}\right] \]

This expression corresponds to what would be obtained for an {\it adiabatic}
(isentropic) expansion for radiation (massless
particles) evolving in the cosmological background.

This behavior only appears at {\it leading} order in the high temperature
expansion. There are subleading terms that we computed subsequently \cite{bdh}.
To avoid cluttering of notation, we will set $k_B=\hbar=1$ in what follows.

We define
\begin{equation}
m^2(T_o)\equiv m^2_R+ \xi_R {\cal{R}}(t_o)+ \frac{\lambda_R}{2}
\phi^2(t_o)  + \frac{\lambda_R}{2}\langle \eta^2(t_o) \rangle_R
\end{equation}
and we will assume that $m^2(T_o) \ll T_o^2$.  Since we are interested in
the description of a phase transition, we will write
\begin{equation}
m^2_R+ \xi_R {\cal{R}}(t_o)+ \frac{\lambda_R}{2}\phi^2(t_o)  =
-\frac{\lambda_R T^2_c}{24} ; \; \; \;  T^2_c >0 \label{Tc}
\end{equation}

Thus, to leading order in $T_o$
\begin{equation}
 m^2(T_o) = \frac{\lambda_R}{24}(T_o^2-T^2_c)  \label{mofT}
\end{equation}
Our high temperature expansion will assume {\it fixed} $m(T_o)$ and
$m(T_o)/T_o \ll 1$.

It becomes  convenient to define the variable
\begin{eqnarray}
& & x^2 = \frac{k^2}{T_o^2 a^2(t_o)}+\frac{m^2(T_o)}{T_o^2} \nonumber \\
& & \mid \varphi^H_k(t) \mid^2 = \left|
\varphi^H(a(t_o) \sqrt{x^2T_o^2-m^2(T_o)};t)
\right |^2
 \nonumber
\end{eqnarray}
Recall from our WKB analysis that the leading behavior for
$k\rightarrow \infty$ is (see equation (\ref{divin})
\[ \frac{\mid \varphi^H_k(t) \mid^2}{2} \rightarrow \frac{1}{2a^2(t)k} \]
adding and subtracting this leading term in the integral $J$ and performing
the above change of variables, we have
 \begin{eqnarray}
   J & = & J_1+J_2 \nonumber \\
 J_1 & = & \left[\frac{a(t_o)}{a(t))}\right]^3
\left(\frac{T_o}{\pi}\right)^2 \int^{\infty}_{\frac{m(T_o)}{T_o}}
\frac{x dx}{e^x-1}
\left[a^3(t)\sqrt{x^2T_o^2-m^2(T_o)}
\frac{\mid \varphi^H_k(t) \mid^2}{2}-\frac{a(t)}{2a(t_o)}
\right] \nonumber \\
 J_2 & = & \frac{T_o^2}{2\pi^2}
\left[\frac{a(t_o)}{a(t)}\right]^2
\int^{\infty}_{\frac{m(T_o)}{T_o}}
\frac{x dx}{e^x-1}  \nonumber \\
  & = &  \frac{1}{2}T_o^2 \left[\frac{a(t_o)}{a(t)}\right]^2
\left[\frac{1}{12}-\frac{m(T_o)}{2\pi^2T_o}+\frac{m^2(T_o)}{8\pi^2T_o^2}+
{\cal{O}}\left(\frac{m^3(T_o)}{T_o^3}\right)+\cdots \right] \label{J2}
\end{eqnarray}

We now must study the high temperature expansion of $J_1$. We
restrict ourselves to the determination of the linear and logarithmic
dependence on $T_o$.
For this purpose, it becomes convenient to introduce yet another
change of variables $ x = \frac{m(T_o)}{T_o}z $
and use the fact that in the limit $ T_o \gg m(T_o)$,
\[\frac{z}{e^{\frac{m(T_o)}{T_o}z}-1} \approx \frac{T_o}{m(T_o)}[1-
\frac{m(T_o)}{2T_o}z + \cdots ] \]
This yields the following
linear and logarithmic terms in $T_o$:
\begin{equation}
J_{1lin} = \left[\frac{a(t_o)}{a(t)}\right]^3 \frac{T_o m(T_o)}{\pi^2}
\int^{\infty}_{1} dz \left\{a^3(t) m(T_o)\sqrt{z^2-1}\frac{\mid \varphi^H_k(t)
\mid^2}{2}-
\frac{a(t)}{2a(t_o)} \right\}
\label{linearT}
\end{equation}
Note that the above integral is finite.

The logarithmic contribution is obtained by keeping the
${\cal{O}}(1/k^3)$ in the large momentum expansion of $\mid \varphi^H_k(t)
\mid^2$
given by equation (\ref{divin}) (in terms of the new variable $z$).
We obtain after some straightforward algebra:

\begin{equation}
J_{1log} = -\frac{\ln\left[\frac{m(T_o)}{T_o}\right]}{8\pi^2}
\left\{-\frac{{\cal{R}}}{6}-\frac{\dot{a}^2(t_o)}{a^2(t)}+
\left[\frac{a(t_o)}{a(t)}\right]^2\left[m^2(T_o)+
\frac{\lambda_R T^2_c}{24}\right] -\frac{\lambda_R T^2_c}{24}\right\}
\end{equation}
That is, in the limit $T_o >> m(T_o),~ J_1 = J_{1lin} + J_{1log} +
O((T_o)^0)$.

Comparing the ${\cal{O}}(T_o^2, T_o, \ln T_o)$ contributions it becomes
clear that they have very different time dependences through the
scale factor $a(t)$. Thus the high temperature expansion as presented
will not remain accurate at large times since the term quadratic in $T_o$
may become of the same order or smaller than the linear or logarithmic terms.
The high
temperature expansion and the long time limit are thus not interchangeable,
and any high temperature expansion is thus bound to be valid only within
some time regime that depends on the initial value of the temperature and
the
initial conditions.

As an illustration, we calculated $J_{1lin}$ explicitly in
the case of de Sitter space \cite{bdh}.

\section{Large Time Limit}

The expansion factor $a(t)$ tends to infinity in the limit $t \to
\infty $. As we have just seen, the high temperature expansion breaks
down in this limit. Physical momenta are given by
\begin{equation}
 l = { k \over {a(t)}}
\end{equation}
For large  $a(t)$, only comoving momenta $ k = l\, a(t) \to \infty $
will be relevant. Thus, we can again use the WKB method to evaluate
the mode functions $\varphi^H_k(t)$  in this regime.
Let us consider for example the de Sitter universe ($a(t) = a_o e^{H
t} $). We find from eq.(\ref{harteqofmot})
\begin{equation}
\left|\varphi^H_{ l a(t)}(t) \right|^2 \buildrel{ a(t)\to \infty}\over
= {{H^2 + l^2}\over{2l^2\sqrt{l^2 + {{m(T_o)^2}\over {a(t)^2}}}}}
\left[ 1 + {1 \over {2l^2}}\left( {{m(T_o)^2}\over {a(t)^2}} +
{{{\dot a}(t_0)^2}\over {a(t)^2}} \right) + O \left( {1 \over {[a(t)l]^4}}
 \right)\right]
\label{grat}
\end{equation}
for $ m/H << 1 $. Using the asymptotic behaviour (\ref{grat}) in
eqs.(\ref{etaren})-(\ref{jota}) leads to \cite{bis}
\begin{equation}
\langle \eta^2 \rangle_R  = {1 \over {(2\pi)^2}} \left\{H^2 Z({K \over
{m(T_0)}}) + Y(T_0) \left[{{ a(t_0)}\over {a(t)}}\right]^2 + O(a(t)^{-4})
\right\}
\end{equation}
where
\begin{eqnarray}
 Z({K \over {m(T_0)}}) & = & {1 \over 2}\, (x-1) \left( 1 +
{{{\dot a}(t_0)^2}\over {m^2}} \right) +
\log \left({2 \over {1 + x}}\right)
{\rm with~ } x  \equiv  \sqrt{1 + {{m^2}\over {K^2}}}  \\
Y(T)  & = & \int_0^{\infty} {{k \; dk}\over{e^{{\sqrt{k^2 +
m(T)^2}}\over T}-1}} + {{m^4}\over{4K^2}}{1\over{(1+x)^2}}+
{{{\dot a}(t_0)^2}\over 2}\log\left({2 \over {1 + x}}\right)
\end{eqnarray}
It should be noticed that the integral in $ Y(T) $ is the mean value
of $ {1 \over k} $ for a free Bose gas at temperature $T_0$. Notice
that both  $ Y(T) $ and  $ Z({K \over {m(T_0)}}) $ depend on the
renormalization point $ K $ and that  $Z({K \over {m(T_0)}}) $ is {\bf
positive} for all $ K $ .

Assuming that $ \phi(t) \to \phi_{\infty}$ (a constant) for $ t \to
\infty $, the Hartree equations  (\ref{rennewdiffeqU}) yield for
$ t \to \infty $,
\begin{equation}
 \phi_{\infty}^2 = - {2 \over {\lambda_r}} \left[ m_r^2 + 12 H^2
\xi_r \right] - H^2 \;  Z({K \over {m(T_0)}})
\label{harasi}
\end{equation}
Since  $ Z({K\over{m(T_0)}}) > 0 $
and we consider $\lambda_r <<1 $,
this equation has a solution provided
\begin{equation}
m_r^2 + 12 H^2 \xi_r < 0
\label{roto}
\end{equation}
The physical interpretation is as follows, in the symmetry breaking
case where eq.(\ref{roto}) holds, the order parameter $ \phi(t) $ for
large times $t$ (when the universe cools down fast)
tends to the non-zero value $ \phi_{\infty} $
given by eq.(\ref{harasi}) independent of the initial value
$\phi(t_0)$.

The equal time two-point field correlator is given in the Hartree approximation
by
\begin{equation}
\Delta(\vec{x},t) = \langle \eta(\vec{x},t) \eta(0,t)\rangle =  \frac{1}{2}
 \int \frac{d^3k}{(2\pi)^3} ~ e^{i\vec{k}.\vec{x}}
\mid \varphi^H_k(t) \mid^2 \coth\left[\beta_o\hbar {\cal{W}}_k(t_o)/2
\right], \label{corr}
\end{equation}
For long times [ $ a(t) >> 1 $ ] we find using the previous WKB
results for $\mid \varphi^H_k(t) \mid^2  $ (\ref{divin}),
\begin{eqnarray}
\Delta(\vec{x},t) & \buildrel{ a(t)\to \infty}\over =
& \left({H \over {2 \pi}}\right)^2 \; {{A(r)}\over
r} + {{B(r)}\over{[2\pi r a(t)]^2}}~,~~~~{\rm where} \\
A(r) & = & {{\pi}\over {2m}} - \int_r^{\infty}dr\;K_0(mr)\quad,\quad
B(r) = mr K_1(mr) ~,~~~ r \equiv|\vec{x}|
\end{eqnarray}
We find in terms of the physical (redshifted) length $ R \equiv r\,
a(t) $ when $t \to \infty$ and $R$ is kept fixed:
\begin{equation}
\Delta(\vec{x},t) \buildrel{ a(t)\to \infty}\over=
 {1 \over {(2 \pi R)^2}} + ({H \over
{2\pi}})^2 \left[ Ht - \log(2mR) + 1 - \gamma \right] + O(\log{ a(t)}/[a(t)]^2)
\end{equation}
Notice that $m=m(T_0)$ fixes the scale of the correlations.

\end{document}